\documentclass[12pt]{article}
\usepackage{graphicx}
\begin{document}

\begin{center}

\vbox{\vspace{10mm}}
{\LARGE \bf Dirac Matrices and \\[1ex]
 Feynman's Rest of the Universe}

\vspace{10mm}

Young S. Kim\footnote{electronic mail: yskim@umd.edu} \\
Department of Physics, University of Maryland, College Park,
Maryland 20742 \\

\vspace{5mm}

Marilyn E. Noz \footnote{electronic mail: marilyne.noz@gmail.com} \\
Department of Radiology, New York University, New York, New York
10016

\end{center}

\vspace{10mm}

\abstract{There are two sets of four-by-four matrices introduced
by Dirac. The first set consists of fifteen Majorana matrices derivable
from his four $\gamma$ matrices. These fifteen matrices can also serve
as the generators of the group $SL(4,r)$. The second set consists of
ten generators of the $Sp(4)$ group which Dirac derived from two
coupled harmonic oscillators. It is shown possible to extend the
symmetry of $Sp(4)$ to that of $SL(4,r)$ if the area of the phase
space of one of the oscillators  is allowed to become smaller without
a lower limit.  While there are no restrictions on the size of phase
space in classical mechanics, Feynman's rest of the universe makes
this $Sp(4)$-to-$SL(4,r)$ transition possible. The ten generators are
for the world where quantum mechanics is valid. The remaining five
generators belong to the rest of the universe. It is noted that the
groups $SL(4,r)$ and $Sp(4)$ are locally isomorphic to the Lorentz
groups $O(3,3)$ and $O(3,2)$ respectively. This allows us to interpret
Feynman's rest of the universe in terms of space-time symmetry.}

\vspace{20mm}
\noindent published in the Symmetry {\bf 4}, 626-643 (1912).

\newpage

\section{Introduction}\label{intro}
\par
In 1963, Paul A. M. Dirac published an interesting paper on the coupled
harmonic oscillators~\cite{dir63}.  Using step-up and step-down operators,
Dirac was able to construct ten operators satisfying a closed set of
commutation relations.  He then noted that this set of commutation
relations can also be used as the Lie algebra for the $O(3,2)$ deSitter
group applicable to three space and two time dimensions.  He noted
further that this is the same as the Lie algebra for the four-dimensional
symplectic group $Sp(4)$.
\par
His algebra later became the fundamental mathematical language for two-mode
squeezed states in quantum optics~\cite{yuen76,yurke86,knp91,hkny93}.
Thus, Dirac's ten oscillator matrices play a fundamental role in modern
physics.
\par
In the Wigner phase-space representation, it is possible to write the
Wigner function in terms of two position and two momentum variables.
It was noted that those ten operators of Dirac can be translated into
the operators with these four variables~\cite{knp91,hkn95jmp}, which then
can be written as four-by-four matrices.  There are thus ten four-by-four
matrices.  We shall call them  Dirac's oscillator matrices.  They are
indeed the generators of the symplectic group $Sp(4)$.
\par

We are quite familiar with four Dirac matrices for the Dirac equation,
namely $\gamma_{1},~\gamma_{2},~\gamma_{3},$ and $\gamma_{0}.$  They
all become imaginary in the Majorana representation.

From them we can construct fifteen linearly independent four-by-four
matrices.  It is known that these four-by-four matrices can serve as
the generators of the $SL(4,r)$ group~\cite{hkn95jmp,dlee95}.  It is
also known that this $SL(4,r)$ group is locally isomorphic to the
Lorentz group $O(3,3)$ applicable to the three space and three time
dimensions~\cite{hkn95jmp,dlee95}.
\par
There are now two sets of the four-by-four matrices constructed by Dirac.
The first set consists of his ten oscillator matrices, and there are fifteen
$\gamma$ matrices coming from his Dirac equation.  There is thus a difference
of five matrices.  The question is then whether this difference can be explained
within the framework of the oscillator formalism with tangible physics.
\par
It was noted that his original $O(3,2)$ symmetry can be extended to that
of $O(3,3)$ Lorentz group applicable to the six dimensional space consisting
of three space and three time dimensions.  This requires the inclusion of
non-canonical transformations in classical mechanics~\cite{hkn95jmp}.  These
noncanonical transformations cannot be interpreted in terms of the present
form of quantum mechanics.
\par
On the other hand, we can use this non-canonical effect to illustrate the
concept of Feynman's rest of the universe.  This oscillator system can
serve as two different worlds.  The first oscillator is the world in
which we do quantum mechanics, and the second is for the rest of the
universe.  Our failure to observe the second oscillator results in the
increase in the size of the Wigner phase space thus increasing the
entropy~\cite{hkn99ajp}.
\par
Instead of ignoring the second oscillator, it is of interest to see what
happens to it.  It is shown in this paper that Planck's constant does
not have a lower limit.  This is allowed in classical mechanics, but
not in quantum mechanics.
\par
Indeed, Dirac's ten oscillator matrices explain the quantum world for
the both oscillators.  The set of Dirac's fifteen $\gamma$ matrices
contains his ten oscillator matrices as a subset.  We discuss in this
paper the physics of this difference.

\par
In Sec.~\ref{majora}, we start with Dirac's four $\gamma$ matrices in
the Majorana representation and construct all fifteen four-by-four
matrices applicable to the Majorana form of the Dirac spinors.
Sec.~\ref{dcouple} reproduces Dirac's derivation of the $O(3,2)$
symmetry with ten generators from two coupled oscillators.  This group
is locally isomorphic to $Sp(4)$, which allows canonical transformations
in classical mechanics.
\par
In Sec.~\ref{o33sym}, we translate Dirac's formalism into the language
of the Wigner phase space.  This allows us to extend the $Sp(4)$ symmetry
into the non-canonical region in classical mechanics.  The resulting
symmetry is that of $SL(4,r)$, isomorphic to that of the Lorentz group
$O(3,3)$ with fifteen generators.  This allows us to establish the
correspondence between Dirac's Majorana matrices with those $SL(4,r)$
four-by-four matrices applicable to the two oscillator system, as well
as the fifteen six-by-six matrices which serve as the generators of
the $O(3,3)$ group,
\par
Finally, in Sec.~\ref{restof}, it is shown that the difference between
the ten oscillator matrices and the fifteen Majorana matrix can serve
as an illustrative example of Feynman's rest of the
universe~\cite{hkn99ajp,fey72}.

\section{Dirac Matrices in the Majorana Representation}\label{majora}
 \par
Since all the generators for the two coupled oscillator system can be
written as four-by-four matrices with imaginary elements, it is convenient
to work with Dirac matrices in the Majorana representation, where the
all the elements are imaginary~\cite{dlee95,majora32,itzyk80}
In the Majorana representation, the four Dirac $\gamma$ matrices are
\begin{eqnarray}\label{majo01}
&{}& \gamma_1 = i\pmatrix{\sigma_3 & 0 \cr 0 & \sigma_3}, \qquad
\gamma_2 = \pmatrix{0 &  -\sigma_2 \cr \sigma_2 & 0},    \nonumber \\[1ex]
&{}&  \gamma_3 = -i\pmatrix{\sigma_1 & 0 \cr 0 & \sigma_1}, \qquad
\gamma_0 = \pmatrix{0 & \sigma_2 \cr \sigma_2 & 0},
\end{eqnarray}
where
$$
\sigma_1 = \pmatrix{0 & 1 \cr 1 & 0}, \quad
\sigma_2 = \pmatrix{0 & -i \cr i & 0}, \quad
\sigma_1 = \pmatrix{1 & 0 \cr 0 & -1} .
$$
These $\gamma$ matrices are transformed like four-vectors under Lorentz
transformations.  From these four matrices, we can construct one
pseudo-scalar matrix
\begin{equation} \label{majo02}
\gamma_5 = i\gamma_0 \gamma_1 \gamma_2 \gamma_3 =
     \pmatrix{\sigma_2 &  0 \cr 0 & -\sigma_2} ,
\end{equation}
and a pseudo vector $i\gamma_5 \gamma_{\mu}$ consisting of
\begin{eqnarray}\label{majo03}
&{}&  i\gamma_5 \gamma_1 = i\pmatrix{- \sigma_1 & 0 \cr 0 & \sigma_1}, \qquad
i\gamma_5 \gamma_2 = -i\pmatrix{0 & I \cr I & 0},
\nonumber \\[1ex]
&{}& i\gamma_5\gamma_0 = i\pmatrix{0 & I \cr  -I & 0} , \qquad
i\gamma_5\gamma_3 = i\pmatrix{-\sigma_3 &  0 \cr 0 & +\sigma_3} .
\end{eqnarray}

\par

In addition, we can construct the tensor  of the $\gamma$ as
\begin{equation}\label{maj04}
T_{\mu\nu} = \frac{i}{2}\left(\gamma_{\mu}\gamma_{\nu}
      - \gamma_{\nu}\gamma_{\mu} \right).
\end{equation}
This antisymmetric tensor has six components.  They are
\begin{equation}\label{majo05}
i\gamma_0 \gamma_1 = - i\pmatrix{0 & \sigma_1 \cr \sigma_1 & 0}, \quad
i\gamma_0 \gamma_2 = - i\pmatrix{-I & 0 \cr 0 & I},  \quad
i\gamma_0\gamma_3 = -i\pmatrix{0 & \sigma_3 \cr \sigma_3 & 0} ,
\end{equation}
and
\begin{equation}\label{majo06}
i\gamma_1 \gamma_2 = i\pmatrix{ 0 & -\sigma_1 \cr \sigma_1 & 0}, \quad
i\gamma_2 \gamma_3 = - i\pmatrix{ 0 & -\sigma_3 \cr \sigma_3 & 0}, \quad
i\gamma_3 \gamma_1 =  \pmatrix{ \sigma_2 & 0 \cr 0 & \sigma_2} .
\end{equation}

\par

There are now fifteen linearly independent four-by-four matrices. They
are all traceless, their components are imaginary~\cite{dlee95}.   We
shall call these Dirac's Majorana matrices.
\par
In 1963~\cite{dir63}, Dirac constructed another set of four-by-four
matrices from two coupled harmonic oscillators, within the framework
of quantum mechanics.  He ended up with ten four-by-four matrices.
It is of interest to compare his oscillator matrices and and his fiftteen
Majorana matrices.

\section{Dirac's Coupled Oscillators}\label{dcouple}
\par
In his 1963 paper~\cite{dir63}, Dirac started with the Hamiltonian for
two harmonic oscillators.  It can be written as
\begin{equation}\label{ham01}
H = {1\over 2} \left(p^{2}_{1} + x^{2}_{1}\right)
+ {1 \over 2} \left(p^{2}_{2} + x^{2}_{2} \right).
\end{equation}
The ground-state wave function for this Hamiltonian is
\begin{equation}\label{wf1}
\psi_{0} (x_{1},x_{2}) = \frac {1}{\sqrt{\pi}}
\exp{ \left\{- {1\over 2}(x^{2}_{1} + x^{2}_{2})
\right\} } .
\end{equation}

\par
We can now consider unitary transformations applicable to the
ground-state wave function of Eq.(\ref{wf1}), and Dirac noted that
those unitary transformations are generated by~\cite{dir63}
\begin{eqnarray}\label{hatdag}
L_{1} &=& {1\over 2}\left(a^{\dag }_{1}a_{2} + a^{\dag }_{2}a_{1}
\right) ,\qquad L_{2} = {1\over 2i}\left(a^{\dag }_{1}a_{2} -
a^{\dag }_{2}a_{1}\right) ,  \nonumber \\[3mm]
L_{3} &=& {1\over 2}\left(a^{\dag}_{1}a_{1} -
a^{\dag}_{2}a_{2} \right) , \qquad
S_{3} = {1\over 2}\left(a^{\dag}_{1}a_{1} +
a_{2}a^{\dag}_{2} \right) ,   \nonumber \\[3mm]
K_{1} &=& -{1\over 4}\left(a^{\dag}_{1}a^{\dag}_{1} + a_{1}a_{1} -
a^{\dag}_{2}a^{\dag}_{2} - a_{2}a_{2}\right) ,	 \nonumber \\[3mm]
K_{2} &=& {i\over 4}\left(a^{\dag}_{1}a^{\dag}_{1} - a_{1}a_{1} +
a^{\dag}_{2}a^{\dag}_{2} - a_{2}a_{2}\right) ,	 \nonumber \\[3mm]
K_{3} &=& {1\over 2}\left(a^{\dag}_{1}a^{\dag}_{2} +
a_{1}a_{2}\right) ,   \nonumber \\[3mm]
Q_{1} &=& -{i\over 4}\left(a^{\dag}_{1}a^{\dag}_{1} - a_{1}a_{1} -
a^{\dag}_{2}a^{\dag}_{2} + a_{2}a_{2} \right) ,   \nonumber \\[3mm]
Q_{2} &=& -{1\over 4}\left(a^{\dag}_{1}a^{\dag}_{1} + a_{1}a_{1} +
a^{\dag}_{2}a^{\dag}_{2} + a_{2}a_{2} \right) ,  \nonumber \\[3mm]
Q_{3} &=& {i\over 2}\left(a^{\dag}_{1}a^{\dag}_{2} -
a_{1}a_{2} \right) .
\end{eqnarray}
where $a^{\dag}$ and  $a$ are the step-up and step-down operators
applicable to harmonic oscillator wave functions.  These operators
satisfy the following set of commutation relations.
\begin{eqnarray}\label{alge11}
&{}& [L_{i}, L_{j}] = i\epsilon _{ijk} L_{k} ,\qquad
[L_{i}, K_{j}] = i\epsilon _{ijk} K_{k} , \qquad
[L_{i}, Q_{j}] = i\epsilon _{ijk} Q_{k} , \nonumber\\[1ex]
&{}&
[K_{i}, K_{j}] = [Q_{i}, Q_{j}] = -i\epsilon _{ijk} L_{k} ,  \qquad
[L_{i}, S_{3}] = 0 ,  \nonumber\\[1ex]
&{}&
[K_{i}, Q_{j}] = -i\delta _{ij} S_{3} , \qquad
[K_{i}, S_{3}] =  -iQ_{i} , \qquad [Q_{i}, S_{3}] = iK_{i} .
\end{eqnarray}

\par
Dirac then determined that these commutation relations constitute the
Lie algebra for the $O(3,2)$ deSitter group with ten generators. This
deSitter group is the Lorentz group applicable to three space coordinate
and two time coordinates.  Let us use the notation $(x, y, z, t, s)$,
with $(x, y, z)$ as space coordinates and $(t, s)$ as two time coordinates.
Then the rotation around the $z$ axis is generated by
\begin{equation}
L_{3} = \pmatrix{0 & -i & 0 & 0 & 0 \cr i & 0 & 0 & 0 & 0 \cr
  0 & 0 & 0 & 0 & 0 \cr 0 & 0 & 0 & 0 & 0 \cr 0 & 0 & 0 & 0 & 0 } .
\end{equation}
The generators $L_1$ and $L_2$ can be also be constructed.  The $K_3$
and $Q_3$ will take the form
\begin{equation}
K_{3} = \pmatrix{0 & 0 & 0 & 0 & 0 \cr 0 & 0 & 0 & 0 & 0 \cr
  0 & 0 & 0 & i & 0 \cr 0 & 0 & i & 0 & 0 \cr 0 & 0 & 0 & 0 & 0 } , \qquad
Q_{3} =\pmatrix{0 & 0 & 0 & 0 & 0 \cr 0 & 0 & 0 & 0 & 0 \cr
               0 & 0 & 0 & 0 & i \cr 0 & 0 & 0 & 0 & 0 \cr
               0 & 0 & i & 0 & 0 } .
\end{equation}
From these two matrices, the generators $K_1, K_2, Q_1, Q_2$ can be
constructed.  The generator $S_3$ can be written as
\begin{equation}
S_{3} = \pmatrix{0 & 0 & 0 & 0 & 0 \cr 0 & 0 & 0 & 0 & 0 \cr
  0 & 0 & 0 & 0 & 0 \cr 0 & 0 & 0 & 0 & -i \cr 0 & 0 & 0 & i &  0 } .
\end{equation}
The last five-by-five matrix generates rotations in the two-dimensional
space of $(t, s)$.
\par
In his 1963 paper~\cite{dir63}, Dirac states that the Lie algebra of
Eq.({\ref{alge11}) can serve as the four-dimensional symplectic group
$Sp(4)$.  In order to see this point, let us go to the Wigner phase-space
picture of the coupled oscillators.

\subsection{Wigner Phase-space Representation}
\par
For this two-oscillator system, the Wigner function is defined
as~\cite{knp91,hkn95jmp}
\begin{eqnarray}\label{wigf1}
\lefteqn{W(x_{1},x_{2}; p_{1},p_{2}) = \left({1\over \pi} \right)^{2}
\int \exp \left\{- 2i (p_{1}y_{1} + p_{2}y_{2}) \right\} } \nonumber\\[3mm]
\mbox{ } & \mbox{ } & \mbox{ }
\times \psi^{*}(x_{1} + y_{1}, x_{2} + y_{2})
\psi (x_{1} - y_{1}, x_{2} - y_{2}) dy_{1} dy_{2} .\hspace*{2cm}
\end{eqnarray}
Indeed, the Wigner function is defined over the four-dimensional phase
space of $(x_{1}, p_{1}, x_{2}, p_{2})$ just as in the case of classical
mechanics.  The unitary transformations generated by the operators of
Eq.(\ref{hatdag}) are translated into linear canonical transformations
of the Wigner function~\cite{knp91}.  The canonical transformations are
generated by the differential operators~\cite{knp91}:
\begin{eqnarray}\label{rotphase}
L_{1} &=& +{i\over 2}\left\{\left(x_{1}{\partial \over \partial p_{2}} -
p_{2}{\partial \over \partial x_{1}} \right) +
\left(x_{2}{\partial \over \partial p_{1}} -
p_{1}{\partial \over \partial x_{2}} \right)\right\}, \nonumber \\[3mm]
L_{2} &=& -{i\over 2}\left\{\left(x_{1}{\partial \over \partial x_{2}} -
x_{2}{\partial \over \partial x_{1}}\right) +
\left(p_{1} {\partial \over \partial p_{2}} -
p_{2}{\partial \over \partial p_{1}}\right)\right\} ,\nonumber \\[3mm]
L_{3} &=& +{i\over 2}\left\{\left(x_{1}{\partial \over \partial p_{1}} -
p_{1}{\partial \over \partial x_{1}}\right) -
\left(x_{2}{\partial \over \partial p_{2}} -
p_{2}{\partial \over \partial x_{2}}\right)\right\} , \nonumber \\[3mm]
S_{3} &=& -{i\over 2}\left\{\left(x_{1}{\partial \over \partial p_{1}} -
p_{1}{\partial \over \partial x_{1}}\right) +
\left(x_{2}{\partial \over \partial p_{2}} -
p_{2}{\partial \over \partial x_{2}}\right)\right\} ,
\end{eqnarray}
and
\begin{eqnarray}\label{sqphase}
K_{1} &=& -{i\over 2}\left\{\left( x_{1}{\partial \over \partial p_{1}} +
p_{1}{\partial \over \partial x_{1}} \right) -
\left(x_{2}{\partial \over \partial p_{2}} +
p_{2}{\partial \over \partial x_{2}} \right)\right\}, \nonumber \\[3mm]
K_{2} &=& -{i\over 2}\left\{\left(x_{1}{\partial \over \partial x_{1}} -
p_{1}{\partial \over \partial p_{1}}\right) +
\left(x_{2}{\partial \over \partial x_{2}} -
p_{2}{\partial \over \partial p_{2}}\right)\right\} , \nonumber \\[3mm]
K_{3} &=& +{i\over 2}\left\{\left(x_{1}{\partial \over \partial p_{2}} +
p_{2}{\partial \over \partial x_{1}}\right) +
\left(x_{2}{\partial \over \partial p_{1}} +
p_{1}{\partial \over \partial x_{2}}\right)\right\} , \nonumber \\[3mm]
Q_{1} &=& +{i\over 2}\left\{\left(x_{1}{\partial \over \partial x_{1}} -
p_{1}{\partial \over \partial p_{1}}\right) -
\left(x_{2}{\partial \over \partial x_{2}} -
p_{2}{\partial \over \partial p_{2}}\right)\right\} ,\nonumber \\[3mm]
Q_{2} &=& -{i\over 2}\left\{\left(x_{1}{\partial \over \partial p_{1}} +
p_{1}{\partial \over \partial x_{1}}\right) +
\left(x_{2}{\partial \over \partial p_{2}} +
p_{2}{\partial \over \partial x_{2}}\right)\right\} ,\nonumber \\[3mm]
Q_{3} &=& -{i\over 2}\left\{\left(x_{2}{\partial \over \partial x_{1}} +
x_{1}{\partial \over \partial x_{2}} \right) -
\left(p_{2}{\partial \over \partial p_{1}} + p_{1}{\partial \over
\partial p_{2}}\right)\right\} .
\end{eqnarray}

\subsection{Translation into four-by-four matrices}
\par
For a dynamical system consisting of two pairs of canonical variables
$x_{1}, p_{1}$ and $x_{2}, p_{2}$, we can use the coordinate variables
defined as~\cite{hkn95jmp}
\begin{equation}\label{coord2}
\left(\eta _{1}, \eta _{2}, \eta _{3}, \eta _{4} \right) =
\left(x_{1}, p_{1}, x_{2}, p_{2} \right).
\end{equation}
Then the transformation of the variables from
$\eta _{i}$ to $\xi _{i}$ is canonical if~\cite{gold80,abra78}
\begin{equation}\label{symp2}
M J \tilde{M} = J ,
\end{equation}
where $M$ is a four-by-four matrix defined by
$$
M_{ij} = {\partial \over \partial \eta _{j}}\xi _{i},
$$
\noindent and
\begin{equation}
J = \pmatrix{0 & 1 & 0 & 0 \cr -1 & 0 & 0 & 0 \cr
             0 & 0 & 0 & 1 \cr 0 & 0 & -1 & 0} .
\end{equation}
According to this form of the $J$ matrix, the area of the phase space for
$x_{1}$ and $p_{1}$ variables remains invariant, and the story is the
same for the phase space of $x_{2}$ and $p_{2}.$

\par

we can thren write the generators of the $Sp(4)$ group as
$$
L_{1} = \frac{-1}{2}\pmatrix{0 & \sigma_{2}  \cr \sigma_{2} & 0}, \qquad
L_{2} = \frac{i}{2} \pmatrix{0 & -I \cr I & 0} ,
$$
\begin{equation}\label{eq11}
L_{3} = \frac{1}{2}\pmatrix{-\sigma_{2} & 0 \cr 0 & \sigma_{2}} , \qquad
S_{3} = \frac{1}{2}\pmatrix{\sigma_{2}   & 0\cr 0 & \sigma_{2}} .
\end{equation}
and
$$
K_{1} = \frac{i}{2}\pmatrix{\sigma_{1}  & 0 \cr 0 & -\sigma_{1} }, \quad
K_{2} = \frac{i}{2} \pmatrix{\sigma_{3} & 0 \cr 0 & \sigma_{3} } , \quad
K_{3} = -\frac{i}{2} \pmatrix{0 & \sigma_{1} \cr \sigma_{1} & 0} ,
$$
\noindent and
\begin{equation}\label{eq22}
Q_{1} = \frac{i}{2}\pmatrix{-\sigma_{3} & 0 \cr 0 & \sigma_{3}}, \quad
Q_{2} = \frac{i}{2}\pmatrix{\sigma_{1} & 0 \cr 0 & \sigma_{1} } , \quad
Q_{3} = \frac{i}{2}\pmatrix{0 &  \sigma_{3} \cr \sigma_{3}  & 0} .
\end{equation}

These four-by-four matrices satisfy the commutation relations given in
Eq.(\ref{alge11}).  Indeed, the deSitter group O(3,2) is locally isomorphic
to the $Sp(4)$ group.  The remaining question is whether these ten matrices
can serve as the fifteen Dirac matrices given in Sec.~\ref{majora}.  The
answer is clearly No.  How can ten matrices describe fifteen  matrices?
We should therefore add five more matrices.

\section{Extension to O(3,3) Symmetry}\label{o33sym}
\par
Unlike the case of the Schr\"odinger picture, it is possible to add five
noncanonical generators to the above list.  They are
\begin{eqnarray}\label{rotphase2}
S_{1} &=& +{i\over 2}\left\{\left(x_{1}{\partial \over \partial x_{2}} -
x_{2}{\partial \over \partial x_{1}} \right) -
\left(p_{1}{\partial \over \partial p_{2}} -
p_{2}{\partial \over \partial p_{1}} \right)\right\}, \nonumber \\[3mm]
S_{2} &=& -{i\over 2}\left\{\left(x_{1}{\partial \over \partial p_{2}} -
p_{2}{\partial \over \partial x_{1}}\right) +
\left(x_{2} {\partial \over \partial p_{1}} -
p_{1}{\partial \over \partial x_{2}}\right)\right\} ,
\end{eqnarray}
as well as three additional squeeze operators:
\begin{eqnarray}\label{sqphase2}
G_{1} &=& -{i\over 2}\left\{\left( x_{1}{\partial \over \partial x_{2}} +
x_{2}{\partial \over \partial x_{1}} \right) +
\left(p_{1}{\partial \over \partial p_{2}} +
p_{2}{\partial \over \partial p_{1}} \right)\right\}, \nonumber \\[3mm]
G_{2} &=& {i\over 2}\left\{\left(x_{1}{\partial \over \partial p_{2}} +
p_{2}{\partial \over \partial x_{1}}\right) -
\left(x_{2}{\partial \over \partial p_{1}} +
p_{1}{\partial \over \partial x_{2}}\right)\right\} , \nonumber \\[3mm]
G_{3} &=& -{i\over 2}\left\{\left(x_{1}{\partial \over \partial x_{1}} +
p_{1}{\partial \over \partial p_{1}}\right) +
\left(x_{2}{\partial \over \partial p_{1}} +
p_{1}{\partial \over \partial x_{2}}\right)\right\}.
\end{eqnarray}
These five generators perform  well-defined operations on the Wigner
function.  However, the question is whether these additional generators
are acceptable in the present form of quantum mechanics.

\par
In order to answer this question, let us note that the uncertainty
principle in the phase-space picture of quantum mechanics is stated in
terms of the minimum area in phase space for a given pair of conjugate
variables.  The minimum area is determined by Planck's constant.  Thus we
are allowed to expand  phase space, but are not allowed to contract it.
With this point in mind, let us go back to $G_{3}$ of Eq.(\ref{sqphase2}),
which generates transformations which simultaneously expand one phase
space and contract the other.  Thus, the $G_{3}$ generator is not
acceptable in quantum mechanics even though it generates well-defined
mathematical transformations of the Wigner function.

\par
If the five generators of Eq.(\ref{rotphase2}) and Eq.(\ref{sqphase2})
are added to the ten generators given in Eq.(\ref{rotphase}) and
Eq.(\ref{sqphase}), there are fifteen generators.  They satisfy the
following set of commutation relations.
\begin{eqnarray}\label{o33gen}
&{}& [L_{i}, L_{j}] = i\epsilon _{ijk} L_{k}, \qquad
[S_{i}, S_{j}] = i\epsilon_{ijk} S_{k}, \qquad
                           [L_{i}, S_{j}] = 0,  \nonumber \\[1ex]
&{}&
[L_{i}, K_{j}] = i\epsilon_{ijk} K_{k}, \qquad
[L_{i}, Q_{j}] = i\epsilon_{ijk} Q_{k}, \qquad
[L_{i}, G_{j}] = i\epsilon_{ijk} G_{k},            \nonumber \\[1ex]
&{}&
[K_{i}, K_{j}] = [Q_{i}, Q_{j}] = [Q_{i},
    Q_{j}] = -i\epsilon _{ijk} L_{k} ,   \nonumber \\[1ex]
&{}&
[K_{i}, Q_{j}] = -i\delta_{ij} S_{3} , \qquad
[Q_{i}, G_{j}] = -i\delta_{ij} S_{1} , \qquad
[G_{i}, K_{j}] = -i\delta_{ij} S_{2} ,   \nonumber \\[1ex]
&{}&
[K_{i}, S_{3}] = -iQ_{i} , \qquad [Q_{i}, S_{3}] = iK_{i} ,\qquad
[G_{i}, S_{3}] = 0 ,   \nonumber \\[1ex]
&{}&
[K_{i}, S_{1}] = 0 , \qquad [Q_{i}, S_{1}] = -iG_{i} ,\qquad
[G_{i}, S_{1}] = iQ_{i} ,   \nonumber \\[1ex]
&{}&
[K_{i}, S_{2}] = iG_{i} , \qquad [Q_{i}, S_{2}] = 0 ,\qquad
[G_{i}, S_{2}] = -iK_{i} .
\end{eqnarray}
\par
As we shall see in Subsec.~\ref{isomor}, this set of commutation relations
serves as the Lie algebra for the group $SL(4,r)$ and also for the $O(3,3)$
Lorentz group.
\par
These fifteen four-by-four matrices are written in terms of Dirac's fifteen
Majorana matrices, and are tabulated in Table~\ref{table11}.  There are
six anti-symmetric and nine symmetric matrices.  These anti-symmetric
matrices were divided into two sets of three rotation generators in the
four-dimensional phase space.  The nine symmetric matrices can be divided
into three set of three squeeze generators.  However, this classication
scheme is easier to understand in terms the group $O(3,3)$, discussed
in Subsec.~\ref{isomor}.

\par
\begin{table}[ht]
\caption{$SL(4,r)$ and Dirac matrices. Two sets of rotation generators and
three sets of boost generators. There are 15 generators.}\label{table11}
\vspace{2mm}
\begin{center}
\begin{tabular}{lcccc}
\hline
\hline \\
{} &  {}&    First component &  Second component & Third component \\[1.0ex]
\hline\\
Rotation  & {}&  $L_1 = \frac{-i}{2}\gamma_0$   &
          $L_2 = \frac{-i}{2} \gamma_5\gamma_0$  &
          $ L_3 = \frac{-1}{2}\gamma_5 $
\\[3ex]
\hline\\
Rotation  &{}& $S_1 = \frac{i}{2} \gamma_2\gamma_3$   &
          $S_2 = \frac{i}{2} \gamma_1\gamma_2$  &
          $ S_3 = \frac{i}{2} \gamma_3\gamma_1 $
\\[3ex]
\hline\\
Boost  &{}&   $K_1 =  \frac{-i}{2}\gamma_5\gamma_1$   &
             $K_2 = \frac{1}{2} \gamma_1$  &
             $  K_3 =\frac{i}{2} \gamma_0\gamma_1 $
\\[3ex]
\hline\\
Boost  &{}&   $Q_1 = \frac{i}{2} \gamma_5\gamma_3$   &
              $Q_2 = \frac{-1}{2} \gamma_3$  &
             $ Q_3 = -\frac{i}{2} \gamma_0\gamma_3 $
\\[3ex]
\hline\\
Boost  &{}&   $G_1 = \frac{-i}{2}\gamma_5\gamma_2$   &
               $G_2 = \frac{1}{2} \gamma_2$  &
               $ G_3 = \frac{i}{2} \gamma_0\gamma_2 $
\\[3ex]
\hline
\hline\\[-0.8ex]
\end{tabular}
\end{center}
\end{table}

\subsection{Non-canonical Transformations in Classical Mechanics}\label{noncan}
\par
In addition to Dirac's ten oscillator matrices, we can consider the matrix
\begin{equation}\label{g3}
G_{3} = {i\over 2} \pmatrix{I & 0 \cr 0 & -I} ,
\end{equation}
which will generate a radial expansion of the phase space of the first
oscillator, while contracting that of the second phase space~\cite{kimli89pl},
as illustrated in Fig.~\ref{pspace33}.  What is the physical significance of
this operation?  The expansion of phase space leads to an increase in
uncertainty and entropy~\cite{hkn99ajp,kimli89pl}.

\begin{figure}
\centerline{\includegraphics[scale=0.40]{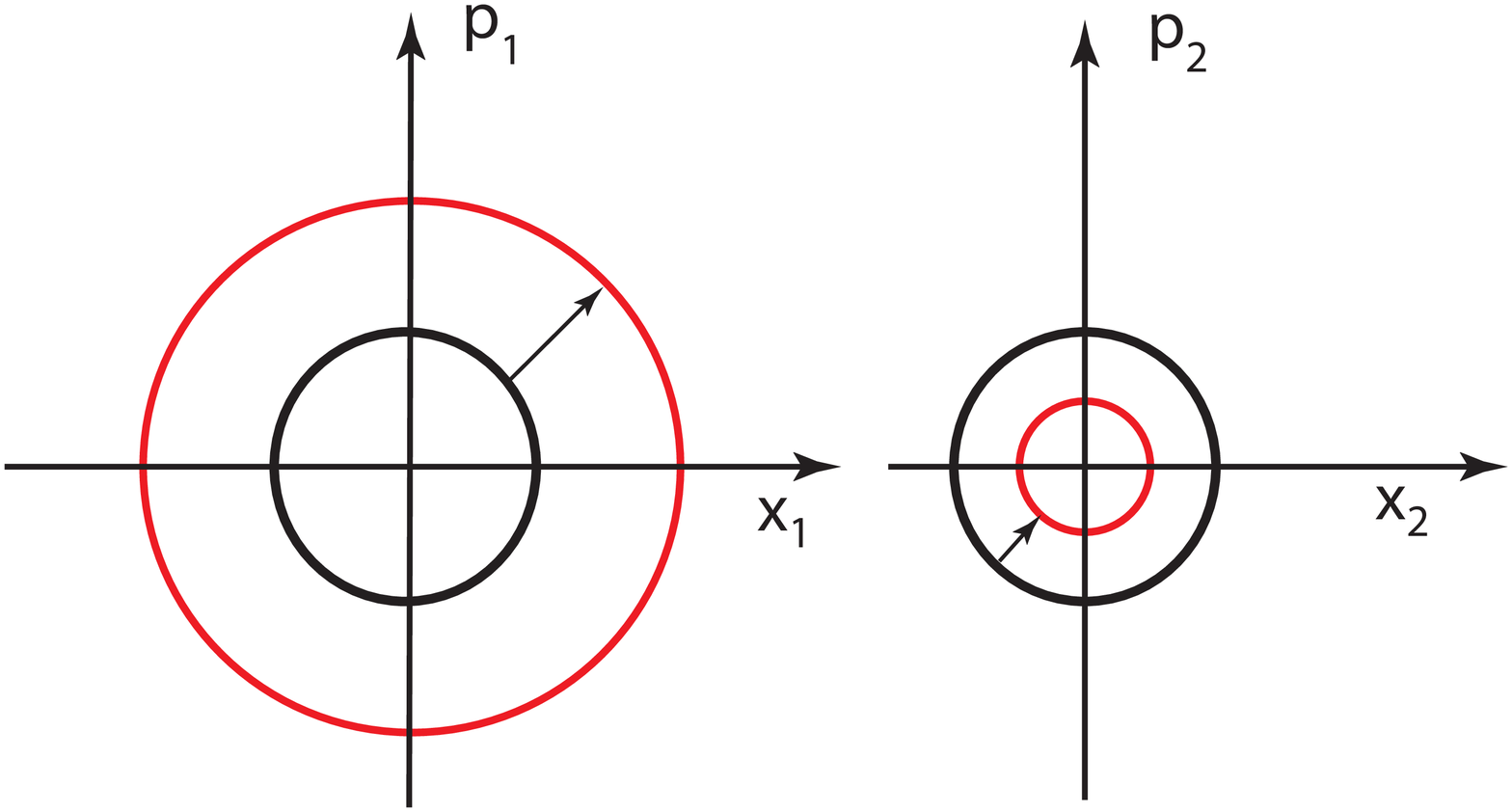}}
\vspace{2mm}
\caption{Expanding and contracting phase spaces.  Canonical transformations
leave the area of each phase space invariant.  Non-canonical
transformations can change them.  Yet the product of these two areas
remain invariant.}\label{pspace33}
\end{figure}
The contraction of the second phase space has a lower limit in quantum
mechanics, namely it cannot become smaller than Planck's constant.
However, there is not such lower limit in classical mechanics.  We shall
go back to this question in
Sec.~\ref{restof}.

\par
In the meantime, let us study what happens when the matrix $G_{3}$ is
introduced into the set of matrices given in Eq.(\ref{eq11}) and
Eq.(\ref{eq22}).  It commutes with $S_{3}, L_{3}, K_{1}, K_{2}, Q_{1}$,
and $Q_{2}$.  However, its commutators with the rest of the matrices
produce four more generators:
\begin{equation}
\left[G_{3}, L_{1}\right] = iG_{2} , \quad
\left[G_{3}, L_{2}\right] = -iG_{1} , \quad
\left[G_{3}, K_{3}\right] = iS_{2} , \quad
\left[G_{3}, Q_{3}\right] = -iS_{1} ,
\end{equation}
where
\begin{eqnarray}\label{eq33}
&{}& G_{1} = {i\over 2}\pmatrix{0 & I \cr I & 0} , \qquad
G_{2} = {1\over 2}\pmatrix{0 & -\sigma_{2} \cr
\sigma_{2} & 0} ,  \nonumber \\[3mm]
&{}& S_{1} = {i\over 2}\pmatrix{0 & \sigma_{3} \cr -\sigma_{3} & 0} , \quad
S_{2} = {i\over 2}\pmatrix{0 & -\sigma_{1} \cr \sigma_{1} & 0} .
\end{eqnarray}
If we take into account the above five generators in addition to the
ten generators of $Sp(4)$, there are fifteen generators.  These generators
satisfy the  set of commutation relations given in Eq.(\ref{o33gen}).
\par
Indeed, the ten $Sp(4)$ generators together with the five new generators
form the Lie algebra for the group $SL(4,r)$.  There are thus fifteeen
four-by-four matrices.  They can be written in terms of the fifteen
Majorana matrices, as given in Table~\ref{table11}.

\par
\begin{table}
\caption{Three-by-three matrices constituting the two-by-two
representation of generators of the $O(3,3)$ group.}\label{table22}
\vspace{2mm}
\begin{center}
\begin{tabular}{lcccccc}
\hline
\hline \\
{} &  {}&  i = 1 &{}& i = 2 &{}& i = 3 \\[1.0ex]
\hline\\[-1.0ex]
$A_{i}$
 & {}&  $\pmatrix{0 & 0 & 0 \cr 0 & 0 & -i \cr 0 & i & 0}$
 & {}& $\pmatrix{0 & 0 & i \cr 0 & 0 & 0 \cr -i & 0 & 0}$
 & {}& $\pmatrix{0 & -i & 0 \cr i & 0 & 0 \cr 0 & 0 & 0}$
\\[4ex]
\hline\\[-1.0ex]
$B_{i}$
 & {}&  $\pmatrix{i & 0 & 0 \cr 0 & 0 & 0 \cr 0 & 0 & 0}$
 & {}& $\pmatrix{0 & 0 & 0 \cr i & 0 & 0 \cr 0 & 0 & 0}$
 & {}& $\pmatrix{0 & 0 & 0 \cr 0 & 0 & 0 \cr i & 0 & 0}$
\\[4ex]
\hline\\[-1.0ex]
$C_{i}$
 & {}&  $\pmatrix{0 & i & 0 \cr 0 & 0 & 0 \cr 0 & 0 & 0}$
 & {}& $\pmatrix{0 & 0 & 0 \cr 0 & i & 0 \cr 0 & 0 & 0}$
 & {}& $\pmatrix{0 & 0 & 0 \cr 0 & 0 & 0 \cr 0 & i & 0}$
\\[4ex]
\hline\\[-1.0ex]
$D_{i}$
 & {}& $\pmatrix{0 & 0 & i \cr 0 & 0 & 0 \cr 0 & 0 & 0}$
 & {}& $\pmatrix{0 & 0 & 0 \cr 0 & 0 & i \cr 0 & 0 & 0}$
 & {}& $\pmatrix{0 & 0 & 0 \cr 0 & 0 & 0 \cr 0 & 0 & i}$
\\[4ex]
\hline
\hline\\[-0.8ex]
\end{tabular}
\end{center}
\end{table}
\par

\subsection{Local Isomorphism  between O(3,3) and SL(4,r)}\label{isomor}
\par
It is now possible to write fifteen six-by-six matrices which generate
Lorentz transformations on the three space coordinates and three time
coordinates~\cite{hkn95jmp}.  However, those matrices are difficult to
handle and do not show existing regularities.  In this section, we write
those matrices as two-by-two matrices of three-by-three matrices.
 \par
For this purpose, we construct four sets of three-by-three matrices
given in Table~\ref{table22}.  There are two sets of rotation generators
\begin{equation}
L_{i} = \pmatrix{A_{i} & 0 \cr 0 & 0}, \qquad
              S_{i} = \pmatrix{0 & 0 \cr 0 & A_{i}},
\end{equation}
applicable to the space and time coordinates respectively.
\par
There are also three sets of boost generators.  In the two-by-two
representation of the matrices given in Table~\ref{table22}, they are
\begin{equation}
K_{i} = \pmatrix{0 & B_{i} \cr \tilde{B_{i}} & 0}, \quad
Q_{i} = \pmatrix{0 & C_{i} \cr \tilde{C_{i}} & 0}, \quad
G_{i} = \pmatrix{0 & D_{i} \cr \tilde{D_{i}} & 0} ,
\end{equation}
where the three-by-three matrices $A_{i}, B_{i}, C_{i},$ and $D_{i}$ are
given in Table~\ref{table22}, and $\tilde{A_{i}}, \tilde{B_{i}},
\tilde{C_{i}}, \tilde{D_{i}}$ are
their transposes respectively.

\par
There is a four-by-four Majorana matrix corresponding to each of hese
fifteen six-by-six matrices, as given in Table~\ref{table11}.

\par

There are of course many interesting subgroups.  The most interesting
case is the $O(3,2)$ subgroup, and there are three of them.  Another
interesting feature in that there are three  time dimensions. Thus,
there are also $O(2,3)$ subgroups applicable to two space and three
time coordinates.  This symmetry between space and time coordinates
could be an interesting future investigation.

\section{Feynman's Rest of the Universe}\label{restof}
 \par
In his book on statistical mechanics~\cite{fey72}, Feynman makes the
following statement.  {\it When we solve a quantum-mechanical problem,
what we really do is divide the universe into two parts - the system in
which we are interested and the rest of the universe.  We then usually
act as if the system in which we are interested comprised the entire
universe.  To motivate the use of density matrices, let us see what
happens when we include the part of the universe outside the system}.
\par

We can use two coupled harmonic oscillators to illustrate what
Feynman says about his rest of the universe.  One of the oscillators
can be used for the world in which we make physical measurements,
while the other belongs to the rest of the universe~\cite{hkn99ajp}.

\par
Let us start with a single oscillator in its ground state.  In
quantum mechanics, there are many kinds of excitations of the
oscillator, and three of them are familiar to us.  First, it can be
excited to a state with a definite energy eigenvalue.  We obtain the
excited-state wave functions by solving the eigenvalue problem for
the Schr\"{o}dinger equation, and this procedure is well known.
\par
Second, the oscillator can go through coherent excitations.  The
ground-state oscillator can be excited to a coherent or squeezed state.
During this process, the minimum uncertainty of the ground state is
preserved.  The coherent or squeezed state is not in an energy
eigenstate.  This kind of excited state plays a central role in
coherent and squeezed states of light which have recently become
a standard item in quantum mechanics.
\par
Third, the oscillator can go through thermal excitations.  This is
not a quantum excitation, but is a statistical ensemble.  We cannot
express a thermally excited state by making linear combinations of
wave functions.  We should treat this as a canonical ensemble.  In
order to deal with this thermal state, we need a density matrix.
\par
For the thermally excited single-oscillator state, the density
matrix takes the form~\cite{fey72,landau58,davies75}.
\begin{equation}\label{den11}
\rho(x, y) = \left(1 - e^{-1/T}\right)
        \sum_{k} e^{-k/T} \phi_{k}(x)\phi_{k}^*(x) ,
\end{equation}
where the absolute temperature $T$ is measured in the scale of
Boltzmann's constant, and $\phi_k(x)$ is the k-th excited state wave
oscillator wave function.  The index ranges from $0$ to $\infty$.
\par
We also use Wigner functions to deal with statistical problems in
quantum mechanics.  The Wigner function for this thermally excited
state is~\cite{fey72,knp91,davies75}
\begin{equation}
W_{T}(x,p) = \frac{1}{\pi} \int e^{-2ipz} \rho(x - z, x + z) dz,
\end{equation}
which becomes
\begin{equation}\label{wf03}
W_{T} = \left[\frac{\tanh(1/2T)}{\pi}\right]
   \exp{\left[-\left(x^2 + p^2\right)\tanh(1/2T)\right]} .
\end{equation}
This Wigner function becomes
\begin{equation}\label{wigf02}
W_{0} = \frac{1}{\pi}
   \exp{\left[-\left(x^2 + p^2\right)\right]} ,
\end{equation}
when $T = 0$.  As the temperature increases, the radius of this
Gaussian form increases from one to~\cite{kimli89pl}
\begin{equation}
\frac{1}{\sqrt{\tanh(1/2T)}}.
\end{equation}

\par
The question is whether we can derive this expanding Wigner function
from the concept of Feynman's rest of the universe.  In their 1999
paper~\cite{hkn99ajp}, Han {\it et al.} used two coupled harmonic
oscillators to illustrate what Feynman said about his rest of the
universe.  One of their two oscillators is for the world in which
we do quantum mechanics and the other is for the rest of the universe.
However, these authors did not use canonical transformations.  In
Subsec.~\ref{cano}, we summarize the main point of their  paper
using the language of canonical transformations developed in the
present paper.
\par
Their work was motivated by the papers by Yurke {\it et al.}~\cite{yupo87}
and by Ekert {\it et al.}~\cite{ekn89}, and  the Barnett-Phoenix version
of information theory~\cite{baph91}.  These authors asked the question of
what happens when one of the photons is not observed in the two-mode
squeezed state.

\par
In Subsec.~\ref{nocan}, we introduce another form of Feynman's rest of the
universe, based on non-canonical transformations discussed in the present
paper.  For a two-oscillator system, we can define a single-oscillator
Wigner function for each oscillator.  Then non-canonical transformations
allow one Wigner function to expand while forcing the other to shrink.
The shrinking Wigner function has  a  lower limit in quantum mechanics,
while there is none in classical mechanics.  Thus, Feynman's rest of the
universe consists of classical mechanics where Planck's constant has no
lower limit.
\par
In Subsec.~\ref{entro}, we translate the mathematics of the expanding
Wigner function into the physical language of entropy.

\subsection{Canonical Approach}\label{cano}
\par

\begin{figure}
\centerline{\includegraphics[scale=0.40]{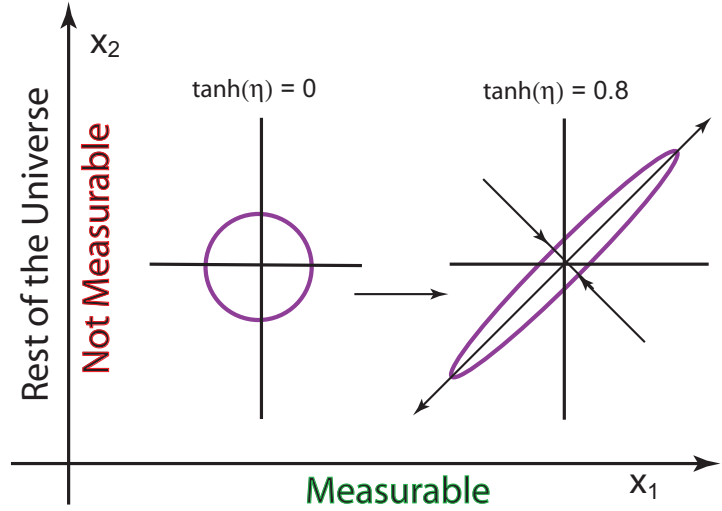}}
\vspace{2mm}
\caption{Two-dimensional Gaussian form for two-coupled oscillators. One
of the variables is observable while the second variable is not observed.
It belongs to Feynman's rest of the universe.}\label{ellipse}
\end{figure}

Let us start with the ground-state wave function for the uncoupled system.
Its Hamiltonian is given in Eq.(\ref{ham01}), and its wave function is
\begin{equation}
\psi_{0}(x_{1},x_{2}) = \frac{1}{\sqrt{\pi}}
\exp{ \left[- \frac{1}{2}\left(x_{1}^{2} + x_{2}^{2}\right) \right] }.
\end{equation}
We can couple these two oscillators by making the following canonical
transformations.  First, let us rotate the coordinate system by $45^o$
to get
\begin{equation}
\frac{1}{\sqrt{2}} \left(x_1 + x_2\right), \qquad
\frac{1}{\sqrt{2}} \left(x_1 - x_2\right),
\end{equation}
Let us then squeeze the coordinate system:
\begin{equation}\label{canx}
\frac{e^{\eta}}{\sqrt{2}} \left(x_1 + x_2\right), \qquad
\frac{e^{-\eta}}{\sqrt{2}} \left(x_1 - x_2\right) .
\end{equation}
Likewise, we can transform the momentum coordinates to
\begin{equation}\label{canp}
\frac{e^{-\eta}}{\sqrt{2}} \left(p_1 + p_2\right), \qquad
\frac{e^{\eta}}{\sqrt{2}} \left(p_1 - p_2\right) .
\end{equation}
\par
Equations (\ref{canx}) and (\ref{canp}) constitute a very familiar
canonical transformation.  The resulting wave function for this
coupled system becomes
\begin{equation}
\psi_{\eta}(x_{1},x_{2}) = {1 \over \sqrt{\pi}}
\exp\left\{- {1\over 4}\left[e^{2\eta}(x_{1} - x_{2})^{2} +
e^{-2\eta}(x_{1} + x_{2} )^{2} \right]\right\} .
\end{equation}
This transformed wave function is illustrated in Fig.~\ref{ellipse}.

\par

As was discussed in the literature for several different
purposes~\cite{knp91,kno79ajp,knp86,gied03}, this wave function can
be expanded as
\begin{equation}\label{expan}
\psi_{\eta }(x_{1},x_{2}) = {1 \over \cosh\eta}\sum^{}_{k}
(\tanh\eta)^{k} \phi_{k}(x_{1}) \phi_{k}(x_{2}) ,
\end{equation}
where the wave function $\phi_{k}\phi(x)$ and the range of summation are
defined in Eq.(\ref{den11}).
From this wave function, we can construct the pure-state density matrix
\begin{equation}
\rho(x_{1},x_{2};x_{1}',x_{2}')
= \psi_{\eta}(x_{1},x_{2})\psi_{\eta}(x_{1}',x_{2}') ,
\end{equation}
which satisfies the condition $\rho^{2} = \rho $:
\begin{equation}
\rho(x_{1},x_{2};x_{1}',x_{2}') =
\int \rho(x_{1},x_{2};x_{1}'',x_{2}'')
\rho(x_{1}'',x_{2}'';x_{1}',x_{2}') dx_{1}'' dx_{2}'' .
\end{equation}
\par
If we are not able to make observations on the $x_{2}$, we should
take the trace of the $\rho$ matrix with respect to the $x_{2}$
variable.  Then the resulting density matrix is
\begin{equation}\label{integ}
\rho(x, x') = \int \psi_{\eta}(x,x_{2})
\left\{\psi_{\eta}(x',x_{2})\right\}^{*} dx_{2} .
\end{equation}
Here, we have replaced $x_{1}$ and $x'_{1}$ by $x$ and $x'$
respectively.  If we complete the integration over the $x_{2}$ variable,
\begin{equation}\label{den22}
\rho(x, x') = \left({1\over \pi\cosh(2\eta)}\right)^{1/2}
\exp{\left\{-\left[{(x + x')^{2} + (x - x')^{2}\cosh^{2}(2\eta) \over
4 \cosh(2\eta)}\right] \right\}} .
\end{equation}
The diagonal elements of the above density matrix are
\begin{equation}
\rho(x, x) = \left({1\over \pi \cosh(2\eta)} \right)^{1/2}
\exp{\left(\frac{-x^{2}}{\cosh(2\eta)} \right)} .
\end{equation}
With this expression, we can confirm the property of the density
matrix: $Tr(\rho) = 1$.
\par
As for the trace of $\rho^{2}$, we can perform the integration
\begin{equation}\label{trace2}
Tr\left(\rho^{2} \right) = \int \rho(x, x') \rho(x',x) dx'dx =
\frac{1}{\cosh(2\eta)} ,
\end{equation}
which is less than one for nonzero values of $\eta$.
\par
The density matrix can also be calculated from the expansion of the
wave function given in Eq.(\ref{expan}).  If we perform the integral
of Eq.(\ref{integ}), the result is
\begin{equation}\label{dmat}
\rho(x,x') = \left({1 \over \cosh\eta}\right)^{2}
\sum^{}_{k} (\tanh\eta)^{2k}
\phi_{k}(x)\phi^{*}_{k}(x') ,
\end{equation}
which leads to $Tr(\rho) = 1$.  It is also straightforward to compute
the integral for $Tr(\rho^{2})$.  The calculation leads to
\begin{equation}
Tr\left(\rho^{2} \right)
= \left({1 \over \cosh\eta}\right)^{4}
\sum^{}_{k} (\tanh\eta)^{4k} .
\end{equation}
The sum of this series becomes to $[1/\cosh(2\eta)]$, as given in Eq.(\ref{trace2}).

\par

We can approach this problem using the Wigner function.  The Wigner
function for the two oscillator system is~\cite{knp91}
\begin{equation}\label{wf01}
W_{0}\left(x_{1},p_{1}; x_{2},p_{2}\right) = \left(\frac{1}{\pi}\right)^{2}
   \exp{\left[-\left(x_{1}^2 + p_{1}^2 + x_{2}^2 + p_{2}^2\right)\right]} .
\end{equation}
If we pretend not to make measurement on the second oscillator coordinate,
the $x_{2}$ and $p_{2}$ variables have to be integrated out~\cite{hkn99ajp}.
The net result becomes the Wigner function for the first oscillator.
\par
The canonical transformation of Eq.(\ref{canx}) and Eq.(\ref{canp}) changes
this Wigner function to
\begin{eqnarray}\label{wf02}
\lefteqn{W(x_{1},x_{2};p_{1},p_{2}) = \left(\frac{1}{\pi} \right)^{2}
\exp\left\{-{1\over 2}\left[e^{2\eta}(x_{1} - x_{2})^{2} +
e^{-2\eta}(x_{1} + x_{2})^{2}
\right. \right.} \hspace{40mm} \nonumber \\ [1.0ex]
\mbox{ } & \mbox{ } & \mbox{ } \left.\left. + e^{-2\eta }
(p_{1} - p_{2})^{2} + e^{2\eta }(p_{1} + p_{2})^{2} \right] \right\}.
\end{eqnarray}

\par
If we do not observe the second pair of variables, we have to integrate
integrate this function over $x_2$ and $p_2$:
\begin{equation}
W_{\eta}\left(x_1, p_1\right) = \int W(x_{1},x_{2};p_{1},p_{2}) dx_2 dp_{2} ,
\end{equation}
and the evaluation of this integration leads to~\cite{hkn99ajp}
\begin{equation}
W_{\eta}(x, p) =  \frac{1}{\pi \cosh\eta}
\exp{\left[- \left(\frac{x^2 + p^2}{\cosh(2\eta)}\right)\right] },
\end{equation}
where we use $x$ and $p$ for $x_{1}$ and $p_{1}$ respectively.
\par
This Wigner function is of the form given in Eq.(\ref{wf03}) for the thermal
excitation, if we identify the squeeze parameter $\eta$ as~\cite{hkn90pl}
\begin{equation}\label{etat}
\cosh(2\eta) = \frac{1}{\tanh(1/2T)} .
\end{equation}
The failure to make measurement on the second oscillator leads to the radial
expansion of the Wigner phase space as in the case of the thermal excitation.

\par

\subsection{Non-canonical Approach}\label{nocan}
\par
As we noted before, among the fifteen Dirac matrices, ten of them
can be used for canonical transformations in classical mechanics, and
thus in quantum mechanics.  They play a special role in quantum
optics~\cite{yuen76,yurke86,knp91,hkny93}.

\par
The remaining five of them can have their roles if the change in the phase
space area is allowed.  In quantum mechanics, the area can be increased,
but it has a lower limit called Plank's constant.  In classical mechanics,
this constraint does not exist.  The mathematical formalism given in this
paper allows us to study this aspect of the system of coupled oscillators.

\par
Let us choose the following three matrices from those in Eq.(\ref{eq11}) and
Eq.(\ref{eq22}).
\begin{equation}\label{eq77}
S_{3} = {1\over 2}\pmatrix{\sigma_{2} & 0 \cr 0 & \sigma_{2}} , \quad
K_{2} = {i\over 2}\pmatrix{\sigma_{3} & 0 \cr 0 & \sigma_{3}}, \quad
Q_{2} = {i\over 2}\pmatrix{\sigma_{1} & 0 \cr 0 & \sigma_{1}} .
\end{equation}
They satisfy the closed set of commutation relations:
\begin{equation}\label{sp2}
\left[S_{3}, K_{2}\right] = iQ_{2}, \quad
\left[S_{3}, Q_{2}\right] = -iQ_{3}, \quad
\left[K_{2}, Q_{2}\right] = -iS_{3} .
\end{equation}
This is the Lie algebra for the $Sp(2)$ group,  This is the symmetry group
applicable to the single-oscillator phase space~\cite{knp91}, with one
rotation and two squeezes.  These matrices generate the same
transformation for the first and second oscillators.
\par
We can choose three other sets with similar properties.  They are
\begin{equation}
S_{3} = {1\over 2}\pmatrix{\sigma_{2} & 0 \cr 0 & \sigma_{2}} , \quad
Q_{1} = {i\over 2}\pmatrix{\sigma_{3} & 0 \cr 0 & -\sigma_{3}}, \quad
K_{1} = {i\over 2}\pmatrix{\sigma_{1} & 0 \cr 0 & -\sigma_{1}} , \nonumber
\end{equation}
\begin{equation}
L_{3} = {1\over 2}\pmatrix{-\sigma_{2} & 0 \cr 0 & \sigma_{2}} , \quad
K_{2} = {i\over 2}\pmatrix{\sigma_{3} & 0 \cr 0 & \sigma_{3}}, \quad
K_{1} = {i\over 2}\pmatrix{-\sigma_{1} & 0 \cr 0 & \sigma_{1}} , \nonumber
\end{equation}
and
\begin{equation}\label{eq8}
L_{3} = {1\over 2}\pmatrix{-\sigma_{2} & 0 \cr 0 & \sigma_{2}} , \quad
-Q_{2} = {i\over 2}\pmatrix{-\sigma_{3} & 0 \cr 0 & \sigma_{3}}, \quad
Q_{2} = {i\over 2}\pmatrix{\sigma_{1} & 0 \cr 0 & \sigma_{1}} .
\end{equation}
\par
These matrices also satisfy the commutation relations given in Eq.(\ref{sp2}).
In this case, the squeeze transformations take opposite directions
in the second phase space.
\par
Since all these transformations are canonical, they leave the area of each
phase space invariant.  However, let us look at the non-canonical generator
$G_{3}$ of Eq.(\ref{g3}).  It generates the transformation matrix of the
form
\begin{equation}
\pmatrix{e^{\eta} & 0 \cr 0 & e^{-\eta}} .
\end{equation}
If $\eta$ is positive, this matrix expands the first phase space while
contracting the second.  This contraction of the second phase space
is allowed in classical mechanics, but it has a lower limit in quantum
mechanics.

\par
The expansion of the first phase space is exactly like the thermal expansion
resulting from our failure to observe the second oscillator which belongs to
the rest of the universe.  If we expand the system of Dirac's ten oscillator
matrices to the world of his fifteen Majorana matrices, we can expand and
contract the first and second phase spaces without mixing them up.  We can
thus construct a model  where the observed world and the rest of the universe
remain separated.  In the observable world, quantum mechanics remains valid
with thermal excitations.  In the rest of the universe, since the area of
the phase space can become small without lower limit, only classical mechanics
is valid.

\par
During the expansion/contraction process, the product of the areas of the
two phase space remains constant.  This may or may not be an extended
interpretation of the the uncertainty principle, but we choose not to
speculate further on this issue.

\par
Let us turn our attention to the fact that the groups $SL(4,r)$ and $Sp(4)$
are locally isomorphic to $O(3,3)$ and $O(3,2)$ respectively.  This means
that we can do quantum mechanics in one of the $O(3,2)$ subgroups of $O(3,3)$,
as Dirac noted in his 1963 paper~\cite{dir63}.  The remaining generators belong
to Feynman's rest of the universe.

\subsection{Entropy and the Expanding Wigner Phase Space}\label{entro}
\par
We have seen how Feynman's rest of the universe increases the radius
of the Wigner function.  It is important to note that the entropy of
the system also increases.
\par
Let us go back to the density matrix.  The standard way to measure
this ignorance is to calculate the entropy defined
as~\cite{landau58,neu32,fano57,blum81,kiwi90pl}
\begin{equation}
S = - Tr\left(\rho \ln(\rho) \right) ,
\end{equation}
where $S$ is measured in units of Boltzmann's constant.  If we use the
density matrix given in Eq.(\ref{dmat}), the entropy becomes
\begin{equation}\label{entro11}
S = \cosh^{2}\eta\ln\left(\cosh^2\eta\right) -
    \sinh^{2}\eta\ln\left(\sinh^2\eta\right) .
\end{equation}
In order to express this equation in terms of the temperature
variable $T$, we write Eq.(\ref{etat}) as
\begin{equation}\label{etat22}
\cosh(2\eta) = \frac{1 + e^{-1/T}}{1 - e^{-1/T}} ,
\end{equation}
which leads to
\begin{equation}
\cosh^2\eta = \frac{1}{1 - e^{-1/T}} , \qquad
\sinh^2\eta = \frac{e^{-1/T}}{1 - e^{-1/T}} .
\end{equation}
\par
Then the entropy of Eq.(\ref{entro11}) takes the form~\cite{hkn99ajp}
\begin{equation}
S = \left(\frac{1}{T}\right)\left\{\frac{1}{\exp{(1/T)} - 1} \right\}
- \ln\left(1 - e^{-1/T}\right).
\end{equation}
This familiar expression is for the entropy of an oscillator state in
thermal equilibrium.  Thus, for this oscillator system, we can relate
our ignorance of the Feynamn's rest of the universe, measured by
of the coupling parameter $\eta$, to the temperature.

\section{Concluding Remarks}
\par
In this paper, we started with the fifteen four-by-four matrices for the
Majorana representation of the Dirac matrices, and the ten generators of
the $Sp(4)$ group corresponding to Dirac's oscillator matrices.  Their
explicit forms are given in the literature~\cite{hkn95jmp,dlee95}, and
their roles in modern physics are well known~\cite{yurke86,knp91,itzyk80}.
We re-organized them into tables.

\par
The difference between these two representations consists of five matrices.
The physics of this difference is discussed in terms of Feynman's rest
of the universe~\cite{fey72}.  According to Feynman, this universe consists
of the world in which we do quantum mechanics, and the rest of the universe.
In the rest of the universe, our physical laws may or may not be respected.
In the case of coupled oscillators, without the lower limit on Planck's
constant, we can do classical mechanics but not quantum mechanics in the
rest of the universe.

\par
In 1971, Feynman {\it et al.}~\cite{fkr71} published a paper on the oscillator
model of hadrons, where the  proton consists of three quarks linked up by
oscillator springs.  In order to treat this problem, they use a
three-particle symmetry group formulated by Dirac in his book on quantum
mechanics~\cite{dir58,hkn80ajp}.  An interesting problem could be to
see what happens to the two quarks when one of them is not observed.
Another interesetinq question could be to see what happens to one of the
quarks when two of them are not observed.
\par

Finally, we note here that group theory is a very powerful tool in
approaching problems in modern physics.  Different groups can share
the same set of commutation relations for their generators.  Recently,
the group $SL(2,c)$, through its correspondence with the $SO(3,1)$
has been shown to be the underlying language for classical and
modern optics~\cite{knp91,bk12mop}.  In this paper, we exploited the
correspondence between $SL(4,r)$ and $O(3,3)$, as well as the
correspondence between $Sp(4)$ and $O(3,2)$ which was first noted by
Paul A. M. Dirac\cite{dir63}.
\par
There could be more applications of group isomorphisms in the future.
A comprehensive list of those correspondences is given in Gilmore's
book on Lie groups~\cite{gilmore74}.

\end{document}